\documentstyle[aps,prl,epsfig,twocolumn,floats]{revtex}
\begin{document}
\draft
\tighten
\twocolumn[\hsize\textwidth\columnwidth\hsize\csname@twocolumnfalse%
\endcsname

\title{Galactic Constraints on the Sources of Ultra-High Energy Cosmic
Rays}
\author{Abraham Loeb$^\star$ \& Eli Waxman$^\dagger$} 
\address{$\star$ Astronomy Department, Harvard University, 60 Garden
Street, Cambridge, MA 02138, USA; aloeb@cfa.harvard.edu\\ $\dagger$ 
Department of Condenced Matter Physics, Weizmann Institute, Rehovot 76100,
Israel; waxman@wicc.weizmann.ac.il } 
\date{\today} 
\maketitle
\begin{abstract}
We show that if the sources of ultra-high energy cosmic rays (UHECRs) with
energies $E\ge 10^{19}$~eV are associated with galaxies, then the
production of UHECRs must occur in transient events.  Our galaxy is
currently at a dim state in between transients.  The time interval between
transients is $\gtrsim 10^{4.5} \min\{1,(\gamma_{\rm
min}/10^3)^{-0.6}\}$~yr for $\gamma_{\rm min}\lesssim 10^7$, where
$\gamma_{\rm min}$ is the minimum Lorentz factor to which protons are
accelerated in the transients. This constraint is satisfied by $\gamma$-ray
bursts.

\end{abstract}

]

\narrowtext

The origin of the observed cosmic rays at different energies is still
unknown (see \cite{Axford94,Nagano00,Halzen02} for recent reviews). As
illustrated in Fig. 1, the cosmic ray spectrum changes its qualitative
behavior as a function of particle energy; it steepens around $\sim 5\times
10^{15}$~eV (the ``knee'') and flattens around $5\times 10^{18}$~eV (the
``ankle'').  Below $\sim 10^{15}$~eV, the cosmic rays are thought to
originate from Galactic supernovae. The composition is dominated by protons
at the lowest energies, and the fraction of heavy nuclei increases with
energy. The proton fraction at $\sim 10^{15}$~eV is reduced to $\sim15\%$
\cite{Burnett90,Bernlohr98}. At yet higher energies, there is evidence that
the fraction of light nuclei increases, and that the cosmic-ray flux above
$5\times 10^{18}$~eV is again dominated by protons \cite{composition}. This
composition change and the flattening of the spectrum around $10^{19}$~eV
(see Fig. 2) suggest that the flux above and below this energy is dominated
by different sources. Since the Galactic magnetic field can not confine
protons above $10^{19}$~eV, it is believed that the nearly isotropic cosmic
ray flux at $E>5\times 10^{18}$~eV, originates from extragalactic sources.
The small, but statistically significant, enhancement of the flux at
$E<3\times 10^{18}$~eV near the Galactic plane \cite{Bird99,Hayashida99},
suggests a Galactic origin at these lower energies.

The Milky-Way disk shows prominently relative to the extragalactic
background in electromagnetic radiation ranging from radio to X-ray
wavelengths. This is a consequence of the fact that our location within the
Galaxy is not representative of a random point in the universe, where the
observed radiation would be nearly isotropic.  The Galactic prominence must
also apply to UHECRs with energies $\gtrsim 10^{19}$ eV. At these energies
the Galactic magnetic field is unable to isotropize the particle orbits,
and the Galactic disk should be on average brighter than the extragalactic
background by orders of magnitude. Contrary to this expectation, the actual
anisotropy of the observed UHECR is $\lesssim 4\%$ in the direction of the
Galactic disk \cite{Bird99,Hayashida99,Nagano00}.  In this paper, we
therefore argue that any generic galactic production of UHECRs must occur
in transient events with a short duty cycle.  Using existing data for the
spectrum and the confinement time of cosmic rays at lower energies, we set
limits on the event rate and the minimum Lorentz factor of the accelerated
particles in these transients.
\begin{figure}[th]
\centerline{\epsfig{file=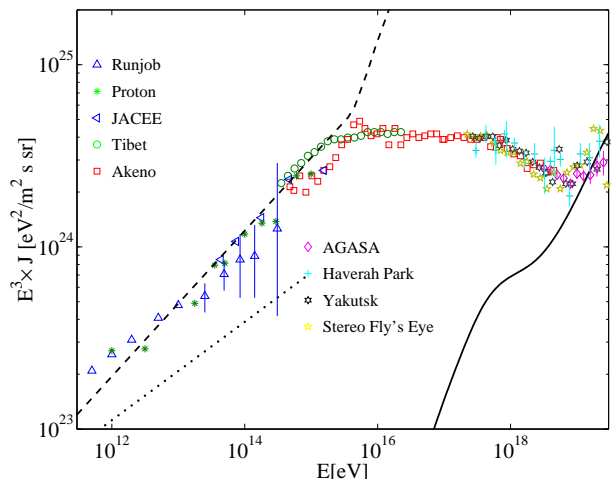, height=2.5in}}
\caption{Data points from [2],
showing the differential flux of all cosmic-ray particles multiplied by
$E^3$.  The dotted line is a power-law fit to the observed proton flux
[3]. The solid line shows the expected volume-averaged, extragalactic
cosmic-ray flux based on the observed UHECR flux (see text and Fig. 2). The
corresponding minimum, time-averaged, Galactic flux is shown as the dashed
line.}
\end{figure}

\paragraph*{The extragalactic cosmic-ray flux.}
In Fig. 2 we compare the UHECR spectrum measured by the three experiments
with the largest current exposure, Yakutsk, Fly's Eye and AGASA, with that
expected from a homogeneous cosmological distribution of sources, each
generating a power-law, differential number flux of high energy protons,
$dn/dE\propto E^{-2}$. This power-law slope is expected for astrophysical
sources which accelerate particles in strong collisionless shocks
\cite{Blandford87,RelShock}, and is found in supernovae \cite{SN} and
$\gamma$-ray bursts \cite{WAG97a}. Experimental differences in the absolute
flux calibration at $3\times10^{18}$~eV yield a systematic over-estimation
of the event energies by $\simeq20\%$ ($\simeq10\%$) in the Yakutsk (AGASA)
experiment as compared to the Fly's Eye experiment; in Fig. 2 we use the
Yakutsk energy normalization.  The calculation of the model flux follows
ref. \cite{W95b}, and assumes a flat universe with $\Omega_{\rm m}=0.3$,
$\Omega_\Lambda=0.7$ and $H_0=65~{\rm km~s^{-1}~Mpc^{-1}}$. The generation
rate of cosmic rays per unit comoving volume is assumed to trace the
redshift evolution of the luminosity density of bright quasars\cite{QSO},
which is also similar to the evolution of the cosmic star formation rate
\cite{SFR}, namely $\dot n_{CR}(z)\propto(1+z)^{\alpha}$ with
$\alpha\approx3$ at low redshifts $z<1.9$, $\dot n_{CR}(z)={\it const}$~
for intermediate redshifts $1.9<z<2.7$, and an exponential decay at high
redshifts $z>2.7$. The cosmic-ray spectrum at energies $> 10^{19}$~eV is
only weakly dependent on the cosmological parameters or the assumed
redshift evolution, because the particles at these energies originate from
distances shorter than a few hundred Mpc.  The spectrum and flux at
$E>10^{19}$~eV is mainly determined by the present ($z=0$) generation rate
and spectrum, which for the model shown in Fig. 2 is given by
\begin{equation}
E^2 \left({d\dot n \over dE}\right)_{z=0}= 0.8\times10^{44}~
{\rm {erg \over Mpc^3~yr}}.
\label{eq:cr_rate}
\end{equation}

\begin{figure}
\centerline{\epsfig{file=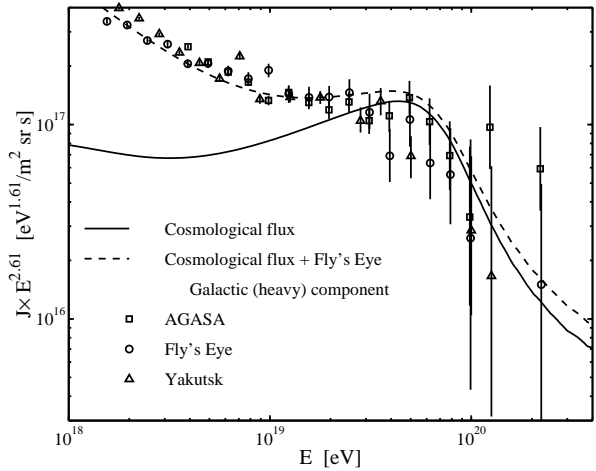, height=2.5in}} 
\caption{ The UHECR flux expected in a cosmological model for which
high-energy protons are produced at the rate given by Eq. (1) (solid line)
compared to the Fly's Eye [9], Yakutsk [10] and AGASA
[11] data (adopted from [12]).  The flux error bars
correspond to $1\sigma$. The highest energy points are derived assuming
that the detected events represent a uniform flux over the energy range
$10^{20}$~eV~--~$3\times10^{20}$~eV.  The dashed line is the sum of the
cosmological model flux and the Fly's Eye fit to the Galactic heavy nuclei
component, $J\propto E^{-3.5}$ [9].  }
\label{fig:fig2}
\end{figure}

The suppression of the flux above $10^{19.7}$~eV is caused by the energy
loss of high energy protons in their interaction with the microwave
background, i.e.  the ``GZK cutoff'' \cite{GZK}. The available data does
not allow to determine the existence (or absence) of the ``cutoff'' with
high confidence. The AGASA results show an excess of events (at a
$\sim2.5\sigma$ confidence level) compared to the model prediction above
$10^{20}{\rm eV}$. This excess is not confirmed, however, by other
experiments; preliminary results from the new HiRes experiment are
consistent with the Fly's Eye data which does not show this excess
\cite{HiRes}.  Since the flux at $10^{20}{\rm eV}$ is dominated by sources
at distances $<100\ {\rm Mpc}$ over which the distribution of known
astrophysical systems (such as galaxies or galaxy clusters) is
inhomogeneous, significant deviations are expected at this energy relative
to the model predictions presented in Fig. 2 under the assumption of a
uniform source distribution \cite{W95b,CR_clustering}.

\paragraph*{Galactic enhancement of observed flux.}
If the production rate of high energy protons given in
Eq. (\ref{eq:cr_rate}) extends to lower energies, then a universal flux of
extragalactic protons would be produced as shown by the solid line in
Fig. 1.  If extragalactic cosmic rays originate in galaxies similar to our
own, then the average Galactic flux would be enhanced relative to this
extragalactic background by two factors. The first factor is cosmological
and is independent of the confinement of cosmic rays by the Galactic
magnetic field.  Ignoring confinement for the moment, we expect that the
extragalactic flux would be diluted relative to the Galactic flux by the
volume filling factor of galaxies $n_{\rm G} V_{\rm G}$ (where $n_{\rm G}$
is the number density of galaxies and $V_{\rm G}$ is the volume of a
galaxy), and be amplified by the ratio between the emission time (which is
of order the Hubble time, $t_{\rm H}$) and the escape time from the
Galactic disk (which is of order the vertical crossing time of the
scale--height of the Galactic disk, $h/c$). The combination of these ratios
yields a Galactic enhancement factor of the cosmic ray flux $J(E)$,
\begin{equation}
f_1={J_{\rm G}\over J_{\rm C}}\approx {1\over c t_{\rm H} n_{\rm G} A_{\rm
G}}\sim 10^3,
\label{eq:enhance1}
\end{equation}
where $A_{\rm G}\equiv V_{\rm G}/h$ is the typical cross-sectional area of
a galactic disk, and the subscripts $G$ and $C$ are used to denote the
Galactic and cosmological values. The numerical value of $f_1\sim 10^3$ is
obtained for the characteristic $A_{\rm G}\sim \pi (5~{\rm kpc})^2$ and
$n_{\rm G}\sim 3\times 10^{-3}~{\rm Mpc^{-3}}$ of $L_\star$ galaxies
similar to the Milky-Way \cite{Madgwick01}.

A second enhancement factor originates from the confinement of cosmic rays
by the Galactic magnetic field. The time that a cosmic ray spends in the
disk is larger than $h/c$ by a factor $f_2$. This factor is directly
determined by observations of daughter nuclei produced in interactions of
cosmic ray nuclei with H and He nuclei in the Galactic disk. The abundance
ratio of daughter to parent nuclei determines the average grammage $\Sigma$
(defined as the path length integral of the gas density) that cosmic ray
nuclei traverse prior to their escape.  The factor $f_2$ is given by the
ratio $\Sigma/\Sigma_{\rm disk}$, where $\Sigma_{\rm disk}=2\times
10^{-3}~{\rm g~cm^{-3}}$ is the surface mass density of the Galactic disk.
For cosmic rays with energies of up to 1~TeV per nucleon \cite{Sigma}
\begin{equation}
f_2\equiv{\Sigma\over \Sigma_{\rm disk}}\approx 10^4 \left({E/Z\over {\rm
GeV}}\right)^{-\alpha},
\label{eq:enhance2}
\end{equation}
where $Z$ is the atomic charge and $\alpha\approx0.6$.  The extrapolation
of this scaling relation to a proton energy $E\sim 10^{16}$~eV gives
$\Sigma\sim \Sigma_{\rm disk}$ or $f_2\approx1$, which implies that the
extrapolation can not be extended to yet higher proton energies since
$\Sigma$ must be larger or equal to $\Sigma_{\rm disk}$. Note that the
extragalactic background is also enhanced by a factor $f_2$ if measured
within the Milky-Way galaxy. Hence the ratio between the local values of
the mean Galactic and extragalactic fluxes remains $f_1$ even in the
presence of confinement.

Figure 1 shows that the observed energy flux $E^2 J(E)$ at $10^{15}$~eV is
larger by a factor of $\sim 10^3$ than its value at $10^{19}$~eV. This
factor is coincidentally comparable to the enhancement factor $f_1$. If, as
commonly assumed, the cosmic rays at $10^{15}$~eV are Galactic in origin
while those at $10^{19}$~eV are extragalactic, then {\it the energy
production rate per galaxy at these two cosmic-ray energies must be
comparable}. Since the particle spectrum expected for acceleration in
astrophysical shocks, $E^2 dN/dE= const$, implies equal amount of energy
per logarithmic energy interval, the above coincidence may seem to suggest
that both the $10^{15}$~eV and $10^{19}$~eV cosmic rays are produced by the
same source population \cite{Milgrom96}.  However, the associated Galactic
component of cosmic rays with energies between $10^{15}$~eV and
$10^{19}$~eV {\it appears to be missing}.  The spectrum steepens instead of
maintaining a $J(E)\propto E^{-2}$ slope in Fig. 1, as expected if the same
sources were producing the observed cosmic-ray flux at $10^{15}$~eV and
$10^{19}$~eV.  The lack of strong anisotropy in the direction of the
Galactic disk at $10^{19}$~eV supports this inference.  {\it Where are the
Galactic counterparts to the extragalactic $10^{19}$~eV cosmic rays?} To
reconcile the data in Fig. 1 with the enhancement factors in
Eqs.~(\ref{eq:enhance1}) and~(\ref{eq:enhance2}), we argue that the UHECRs
must be produced in transient events and that our galaxy is currently at a
dim state in between transients.

\paragraph*{Constraints on transient properties.} The
minimum flux of lower energy cosmic rays produced by the UHECR source
population corresponds to an extrapolation with a spectral slope of
$J(E)\propto E^{-2}$, which is the shallowest slope possible in shock
acceleration (analysis of the observed spectrum at $E>10^{19}$~eV implies
that the source spectral slope can not be shallower than $J(E)\propto
E^{-1.8}$ \cite{W95b}).  The dashed line in Fig. 1 shows this extrapolation
together with the expected enhancement factors $f_1$ and $f_2$ from
Eqs.~(\ref{eq:enhance1}) and~(\ref{eq:enhance2}), assuming conservatively
that Eq. (\ref{eq:enhance2}) with $\alpha=0.6$ holds for protons up to
$E\sim 10^{16}$~eV. At energies exceeding $\sim1$~TeV, the expected mean
Galactic flux of protons exceeds the observed value. The fact that the bulk
of the Galactic cosmic rays in this energy range are missing is supported
by the isotropy of the observed cosmic rays at $10^{19}$~eV.  The Galactic
field is unable to isotropize the distribution to the observed level since
the Larmor radius of a particle of atomic charge $Z$ in the Galactic field
$B$ is $\sim 10~{\rm kpc}(E/10^{19}~{\rm eV})/(B/\mu{\rm G})$,
significantly larger than the Galactic scale height of this magnetic field.

Our assumption that Eq. (\ref{eq:enhance2}) holds for protons up to $E\sim
10^{16}$~eV is conservative since it implies that $\Sigma=\Sigma_{\rm
disk}$ at $E\sim 10^{16}$~eV. If $\Sigma$ is higher, e.g. due to a smaller
value of $\alpha$ beyond 1~TeV per nucleon, then the expected Galactic
counterpart to the extragalactic flux will be higher than shown in Fig. 1.

The confinement time of cosmic rays with an energy of $\sim1$~GeV per
nucleon is measured to be $\sim 10^{7.5}$~yr, based on the survival
fraction of radioactive cosmic-ray nuclei \cite{Webber98}. As the
cosmic-ray energy increases, the decline in $\Sigma$ described in
Eq. (\ref{eq:enhance2}) may be caused either by a decrease in the
confinement time or by a decrease in the time-averaged gas density through
which the cosmic rays propagate (e.g., due to an increase in the thickness
$h_{\rm CR}$ of the disk in which the cosmic rays are confined to a value
larger than the thickness $h$ of the Galactic gaseous disk).  Assuming,
conservatively, that the decrease in $\Sigma$ is only due to a decrease in
the confinement time, we get
\begin{equation}
\tau_{\rm conf}\approx 10^{7.5}~{\rm yr}
\left({E/Z\over {\rm GeV}}\right)^{-\alpha}.
\end{equation}
In order that the Galactic counterpart to the extragalactic $\sim
10^{19}$~eV cosmic ray protons will not exceed the observed proton flux
above $\sim1$~TeV, the time between Galactic transients must satisfy
$\tau_{\rm trans}> f_{\rm vis}\times \min\{\tau_{\rm conf}(E=\gamma_{\rm
min}{\rm GeV}), \tau_{\rm conf}(E=1~{\rm TeV})\}$. Here, $\gamma_{\rm min}$
is the minimum Lorentz factor of the accelerated protons and $f_{\rm vis}$
is the fraction of all Galactic transients which are visible to us. Since
diffusion perpendicular to the Galactic disk drains the cosmic-ray flux
from a distant source, a transient would be visible to us only if it has
occurred within a distance comparable to the scale height $h_{\rm CR}$ of
the cosmic-ray disk.  The estimated scale-height, based on the
time-averaged density through which the cosmic rays propagate, is $\sim 3$
kpc \cite{Webber98} implying $f_{\rm vis}\sim 0.1$.  We therefore obtain
\begin{equation}
\tau_{\rm trans}\gtrsim 10^{4.5} \left({f_{\rm vis}\over 0.1}\right)
\min\left[\left({\gamma_{\rm min}\over10^3}\right)^{-0.6},1\right]~{\rm
yr},
\label{eq:t_burst}
\end{equation} 
as long as $\gamma_{\rm min}\lesssim 10^7$. At higher values of
$\gamma_{\rm min}$ for which $\Sigma\sim \Sigma_{\rm disk}$ and the cosmic
rays escape freely from the Galactic disk, the only constraints are that
$\tau_{\rm trans}$ must be longer than the exposure time of modern
experiments (several decades) and that the transients have a short duty
cycle.

Radioactive dating provides only a conservative lower limit on the
confinement time due to the potential existence of an extended cosmic-ray
halo.  We note that the above conclusions remain unchanged even if the
$10^{19}$~eV cosmic rays are heavy nuclei since the factor $f_1$ is
independent of composition.

\paragraph*{Discussion.}
Among the most likely astrophysical sources for the unsteady production of
UHECRs are Gamma-Ray Bursts (GRBs) and active galactic nuclei (AGNs).
While AGNs (such as SgrA* in the Milky Way galaxy) are known to be
transient just as GRBs, the transient properties of GRBs are
better-determined and allow a more detailed comparison with the required
properties of UHECR sources. The GRB rate per comoving volume today is
$\sim 0.5\times 10^{-9}~{\rm Mpc^{-3}~yr^{-1}}$ \cite{Schmidt01}.  This
rate is derived based on the observed value at a redshift $z\sim1$,
assuming that the rate follows the redshift evolution of the cosmic star
formation rate and that the emission of $\gamma$--rays by the bursts is
isotropic.  If the emission is confined to a double sided jet with a small
angular radius $\theta$, then the burst rate is increased by a factor
$4\pi/2\pi\theta^2$.  GRB observations indicate a characteristic opening
angle $\theta\gtrsim0.1$ \cite{Freedman01,Frail01}.  The implied temporal
separation between bursts in $L_\star$ galaxies is thus $\tau_{\rm GRB}\sim
10^{4.5}(\theta/0.1)^2$~yr, consistent with the constraint in
Eq.~(\ref{eq:t_burst}).

Under the assumption that supernovae dominate the cosmic-ray production
below $10^{15}$~eV, the value of $\gamma_{\rm min}$ for GRBs with
$\tau_{\rm GRB}\sim10^{4.5}$~yr cannot be substantially lower than $\sim
10^3$. This constraint is naturally satisfied in GRBs since the plasma
within which particles are accelerated, is inferred to expand with a bulk
Lorentz factor $\Gamma\sim10^{2.5}$ (e.g. \cite{Dermer01}).

The required energy production rate at $10^{19}$~eV is $\sim 10^{44}~{\rm
erg~Mpc^{-3}~yr^{-1}}$ \cite{W95b,W01Rev}, implying that the cosmic-ray
energy per burst is $10^{51}(\tau_{\rm burst}/10^{4.5}~{\rm yr})~{\rm
ergs}$.  This energy release is surprisingly close to the energy carried by
$\gamma$-rays in GRBs for $\tau_{\rm trans}=10^{4.5}~{\rm yr}$
(corresponding to $\theta\sim 0.1=6^\circ$) \cite{Frail01}, implying that
the local energy generation rate of $\gamma$-rays in GRBs is similar to the
UHECR energy generation rate \cite{W95}\footnote{The AGASA excess above
$10^{20}$~eV has been used to argue that GRBs can not produce the observed
cosmic rays above $10^{20}$~eV \cite{Stecker00}. As explained above, the
excess is not seen in the Fly's Eye, the Yakutsk and the HiRes data, and if
real is likely associated with source inhomogeneities.}. This similarity,
the agreement with the transient rate constraint, and the fact that protons
may be accelerated to energies $>10^{20}$~eV by collisionless internal
shocks of GRBs \cite{W95} support the suggested association between GRBs
and UHECR sources \cite{W95,Vietri95}.
%

The shocks produced due to converging flows during the formation of large
scale structure in the intergalactic medium are also expected to accelerate
cosmic rays \cite{Loeb00}. For the characteristic
magnetic field strength and scale of these shocks, we estimate that the
maximum proton energy is $\sim 10^{17}~{\rm eV}$ and that the intergalactic
cosmic rays may approach the observed flux around this energy but fall
significantly below the observed flux at lower energies.

In closing, we emphasize that the Earth was exposed to a time-averaged
UHECR flux that is greater by three orders of magnitude than the currently
measured value.  During transients the enhancement is much
larger\footnote{The transient event rate and intensity might depend on
particle energy. For beamed GRBs, the bursts at $>10^{19}~{\rm eV}$ will be
less frequent but more intense than those at much lower particle energies,
since the collimation of UHECRs at the source will not be entirely smeared
by the Galactic magnetic field.  When averaged over long timescales the
mean flux is independent of beaming; for a fixed energy output the observed
flux increases due to beaming but the event rate decreases by the inverse
of the same factor.}.  Any fossil record of this intense cosmic-ray
bombardment (e.g. through the detection of stable or radioactive daughter
nuclei in ancient rocks) would provide an important test of our
conclusions.

\paragraph*{Acknowledgments.}
This work was supported in part by grants from the Israel-US BSF
(BSF-9800343), NASA (NAG 5-7039, 5-7768), and NSF (AST-9900877,
AST-0071019). AL acknowledges support from the Einstein Center during his
visit to the Weizmann Institute.

\end{document}